\documentclass{article}
\usepackage[utf8]{inputenc}
\usepackage{amsmath}
\usepackage{graphics}
\usepackage{graphicx}
\usepackage{siunitx}
\usepackage[version=4]{mhchem}
\usepackage[sorting=none]{biblatex}
\usepackage{titling}
\usepackage{authblk}
\author[1,*]{Cameron J. Kopas} 
\author[2]{Dominic P. Goronzy}
\author[2]{Thang Pham}
\author[2]{Carlos G. Torres Castanedo}
\author[2]{Matthew Cheng}
\author[1]{Rory Cochrane}
\author[1]{Patrick Nast}
\author[1]{Ella Lachman}
\author[3,4,5]{Nikolay Z. Zhelev}
\author[6,7]{André Vallières}
\author[7]{Akshay A. Murthy}
\author[8]{Jin-su Oh}
\author[8]{Lin Zhou}
\author[8]{Matthew J. Kramer}
\author[1]{Hilal Cansizoglu}
\author[2,4]{Michael J. Bedzyk}
\author[2,9]{Vinayak P. Dravid}
\author[7]{Alexander Romanenko}
\author[7]{Anna Grassellino}
\author[1]{Josh Y. Mutus}
\author[2,10,11]{Mark C. Hersam}
\author[1]{Kameshwar Yadavalli}
\affil[1]{Rigetti Computing, Berkeley, CA, USA}
\affil[2]{Department of Materials Science and Engineering, Northwestern University, Evanston, IL, USA}
\affil[3]{Center for Applied Physics and Superconducting Technologies, Northwestern University, Evanston, IL, USA}
\affil[4]{Department of Physics and Astronomy, Northwestern University, Evanston, IL, USA}
\affil[5]{Department of Physics, University of Oregon, Eugene, OR, USA}
\affil[6]{Graduate Program in Applied Physics, Northwestern University, Evanston, Illinois 60208, USA}
\affil[7]{Fermi National Accelerator Laboratory, Batavia, IL, USA}
\affil[8]{Ames Laboratory, U.S. Department of Energy, Ames, IA, USA}
\affil[9]{Northwestern University Atomic and Nanoscale Characterization Experimental Center (NUANCE), Northwestern University, Evanston, IL, USA}
\affil[10]{Department of Chemistry, Northwestern University, Evanston, IL, USA}
\affil[11]{Department of Electrical and Computer Engineering, Northwestern University, Evanston, IL, USA}
\affil[*]{Corresponding author: Cameron Kopas, ckopas@rigetti.com}
\date{\today}

\title{Enhanced Superconducting Qubit Performance Through Ammonium Fluoride Etch}

\begin{document}

\maketitle

\section{Abstract}
The performance of superconducting qubits is often limited by dissipation and two-level systems (TLS) losses. The dominant sources of these losses are believed to originate from amorphous materials and defects at interfaces and surfaces, likely as a result of fabrication processes or ambient exposure. Here, we explore a novel wet chemical surface treatment at the Josephson junction-substrate and the substrate-air interfaces by replacing a buffered oxide etch ($\ce{BOE}$) cleaning process with one that uses hydrofluoric acid followed by aqueous ammonium fluoride. We show that the ammonium fluoride etch process results in a statistically significant improvement in median $\text{T}_1$ by $\sim22\%$ ($p=0.002$), and a reduction in the number of strongly-coupled TLS in the tunable frequency range. Microwave resonator measurements on samples treated with the ammonium fluoride etch prior to niobium deposition also show $\sim33\%$ lower TLS-induced loss tangent compared to the BOE treated samples. As the chemical treatment primarily modifies the Josephson junction-substrate interface and substrate-air interface, we perform targeted chemical and structural characterizations to examine materials' differences at these interfaces and identify multiple microscopic changes that could contribute to decreased TLS.

\section{Introduction}

In the growing field of quantum information sciences, quantum computing is a likely successor to state-of-the-art classical high performance computing systems, and efforts to integrate these two computing architectures are already promising \cite{Moller_Vuik_2017, Britt_Humble_2017, Humble_McCaskey_Lyakh_Gowrishankar_Frisch_Monz_2021}. There are several potential quantum computing platforms, including superconducting, ion-trap, photonic, neutral atom, spin and topological approaches \cite{Moses_Baldwin_Allman_Ancona_Ascarrunz_Barnes_Bartolotta_Bjork_Blanchard_Bohn_etal._2023,Arute_Arya_Babbush_Bacon_Bardin_Barends_Biswas_Boixo_Brandao_Buell_etal._2019,Madsen_Laudenbach_Askarani_Rortais_Vincent_Bulmer_Miatto_Neuhaus_Helt_Collins_etal._2022, Graham_Song_Scott_Poole_Phuttitarn_Jooya_Eichler_Jiang_Marra_Grinkemeyer_etal._2022, Veldhorst_Eenink_Yang_Dzurak_2017,Aguado_Kouwenhoven_2020}. Each of these platforms has strengths and trade-offs in terms of performance, reliability, scalability, and complexity. Superconducting qubits are one of the leading approaches that are commonly researched for quantum computing \cite{Ichikawa_2022}. The ability to achieve an addressable quantum system in micron-scale circuits, which can be fabricated by the well understood techniques and processes developed by the CMOS, MEMS, and superconducting electronics industries, as well as the ability to control interactions by purely electronic means have been drivers for the development of the superconducting quantum computing platform over the last two decades \cite{Clarke_Wilhelm_2008}. Furthermore, the use of commonly available fabrication techniques enables the possibility of scaling the qubit count \cite{Vahidpour_OBrien_Whyland_Angeles_Marshall_Scarabelli_Crossman_Yadav_Mohan_Bui_etal._2017, Yost_Schwartz_Mallek_Rosenberg_Stull_Yoder_Calusine_Cook_Das_Day_etal._2020, Foxen_Mutus_Lucero_Graff_Megrant_Chen_Quintana_Burkett_Kelly_Jeffrey_etal._2017, Gold_Paquette_Stockklauser_Reagor_Alam_Bestwick_Didier_Nersisyan_Oruc_Razavi_etal._2021} while being able to correct for device variation using post-processing techniques \cite{Zhang_Srinivasan_Sundaresan_Bogorin_Martin_Hertzberg_Timmerwilke_Pritchett_Yau_Wang_etal._2022, Pappas_Field_Kopas_Howard_Wang_Lachman_Zhou_Oh_Yadavalli_Sete_etal._2024}.

Performance of superconducting qubits is impacted by decoherence through various loss channels. It is well known that dielectric loss from two-level states is a major decoherence source in superconducting qubits \cite{Martinis_Cooper_McDermott_Steffen_Ansmann_Osborn_Cicak_Oh_Pappas_Simmonds_2005}. While the coherence times of superconducting qubits are known to fluctuate unpredictably \cite{Klimov_Kelly_fluctuations_2018}, by systematic analysis and by measuring large numbers of qubits, one can identify the contribution of fabrication related loss channels \cite{Nersisyan_manufacturing_2019}. In typical transmon qubits, the interfaces that affect performance include metal-substrate, substrate-air, metal-air, metal-metal, and dielectrics on the various surfaces and at the Josephson junction barrier.

Several approaches have been investigated to address these loss sources and reduce decoherence in superconducting resonators and qubits. Trenched resonators show improved performance \cite{Vissers_Kline_Gao_Wisbey_Pappas_2012}, due to reduced participation of the substrate-air interface. Furthermore, design changes have been investigated to alter the participation ratio of different interfaces to study their loss contributions \cite{Chu_Axline_Wang_Brecht_Gao_Frunzio_Schoelkopf_2016,Woods_Calusine_Melville_Sevi_Golden_Kim_Rosenberg_Yoder_Oliver_2019}. Additionally, loss in superconducting devices fabricated on Si substrates has been shown to be impacted by surface treatments before and after the qubit electrode deposition \cite{Woods_Calusine_Melville_Sevi_Golden_Kim_Rosenberg_Yoder_Oliver_2019,Quintana_Megrant_Chen_Dunsworth_Chiaro_Barends_Campbell_Chen_Hoi_Jeffrey_etal._2014}. Chemical and/or physical treatments to remove dielectrics from various parts of the device have been shown to help improve device performance partially by removing the native oxide on the Si substrate prior to junction formation \cite{Earnest_Béjanin_McConkey_Peters_Korinek_Yuan_Mariantoni_2018}. Due to the importance of Si substrate preparation for device performance, we looked into testing different pre-cleaning solutions prior to Josephson junction fabrication. Aqueous fluoride solutions are known to provide either a rough surface or a smooth one for Si (100) wafers based on the pH of the solutions \cite{Aldinger_Hines_2012}. Earlier work has shown that an ammonium fluoride etch on a $\ce{Si(111)}$ surface can form an atomically defined and oxidation resistant monohydride surface \cite{Berti_Torres-Castanedo_Goronzy_Bedzyk_Hersam_Kopas_Marshall_Iavarone_2023}. Superconducting Nb films deposited onto these smooth passivated surfaces show improved properties, but it was not known what effect a similar process would have on $\ce{Si(100)}$ surface or on Josephson junctions deposited on such a treated surface.

\begin{figure}
    \centering
    \includegraphics[width=\textwidth]{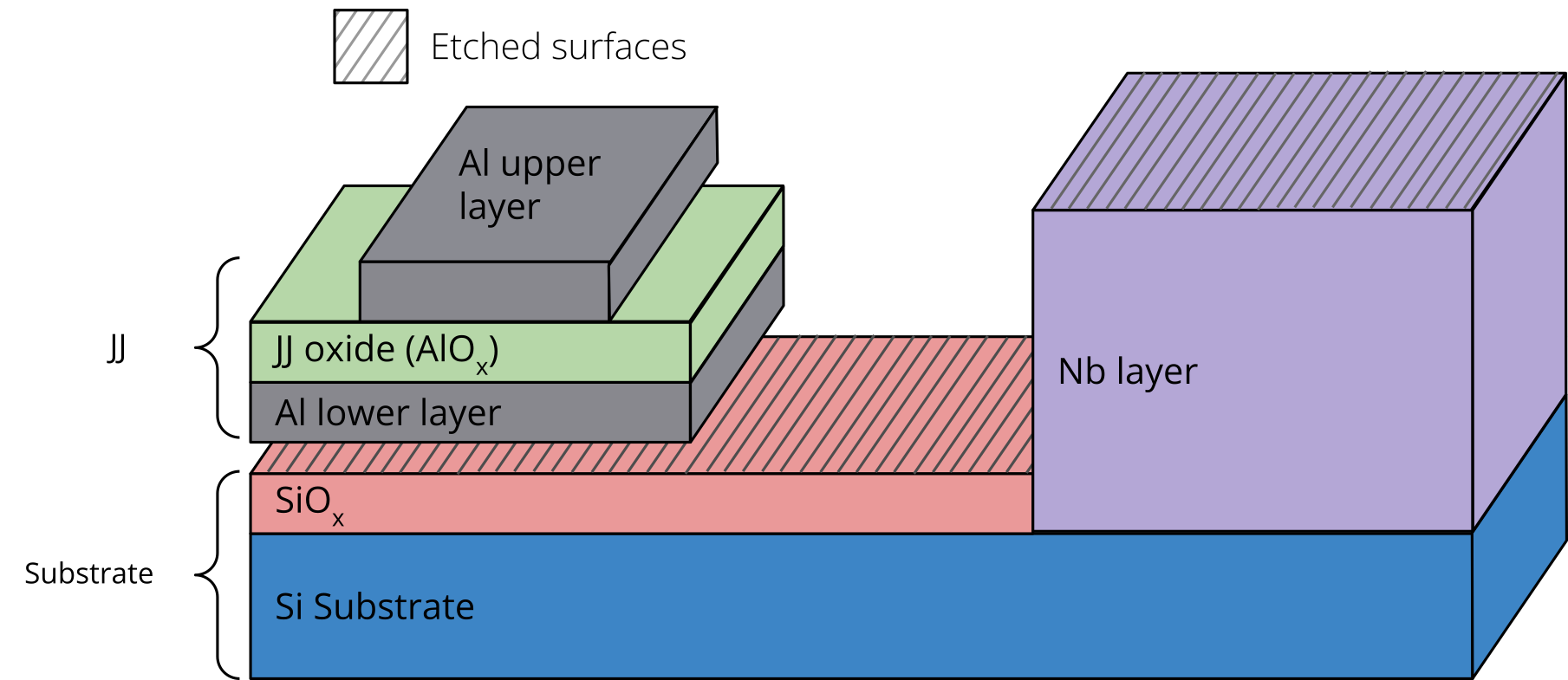}
  \caption{Cross-section schematic of the qubit devices, where the hashed surfaces are those treated during the wet chemical etch processing used in this study. The aluminum does indeed overlap the Nb, but it is not shown for clarity. The $\ce{SiO_x}$ shown is the native oxide formed on a bare $\ce{Si}$ surface, it is included because the oxide removal etch is the focus of this experiment.}
    \label{fig:process}
\end{figure}

In this study we compare two sets of devices fabricated on $\ce{Si(100)}$ wafers that, after dry etching to define the niobium circuit layer fabrication, were prepared using different surface cleaning etches just before the Josephson junction process steps. See Fig. \ref{fig:process} for a schematic of the resulting devices and treated surfaces. The first set (referred to as $\ce{BOE}$ etched from here on) were etched for 2 minutes in a Buffered Oxide Etch (BOE) that is a $5:1$ mixture of $\ce{NH_4F}\ 40\% \ (aq.)\text{:}\ce{HF}\ 49\% \ (aq.)$.
The second set of devices were etched in $\ce{HF}\ 2\%\ (aq.)$ for 1 minute, followed by a deionized (DI) water rinse, and then a 2 minute etch in $\ce{NH_4F}\ 40\% \ (aq.)$, where the wafer was continuously dipped in and out of the $\ce{NH_4F}$ during the etch process for agitation to avoid accumulation of bubbles on the surface. This set will be referred to as $\ce{HF\rightarrow NH_4F}$ from here on. After another DI water rinse, the etched wafers are immediately coated with the bi-layer e-beam lithography resist. After e-beam exposure and pattern development, the junctions are fabricated using a double tilt angle deposition technique. This procedure minimizes the time that the Si and Nb surfaces exposed after resist development are exposed to air. A second lithography and deposition step is used to deposit the patch (or bandage) layer between the Nb circuitry and the junctions \cite{Dunsworth_2017}. More information about the qubit fabrication process and the double-angle lithography process are described in the Supplementary Information section and in Refs. \cite{Nersisyan_manufacturing_2019,Lecocq_CUT_2011}. Throughout this manuscript we use the notation $\ce{HF\rightarrow NH_4F}$ to indicate the $\ce{HF}$ followed by $\ce{NH_4F} $ process steps, where the arrow indicates two sequential processes in time and not a reaction diagram. The sizes of the Josephson junctions used in these test devices range from $0.01\ \mu m^{2}$ to $0.34\ \mu m^{2}$ for each Josephson junction in the SQUID loop. In addition to microwave characterization of resonators and qubits, we performed materials characterization of the affected interfaces using X-ray photoelectron spectroscopy (XPS), time of flight secondary ion mass spectrometry (TOF-SIMS), atomic force microscopy (AFM), contact angle measurements, and scanning transmission electron microscopy (STEM). By combining these materials’ characterizations with extensive device measurements, we demonstrate that an etch process using $\ce{NH_4F} $ reduces TLS and improves dissipation. We also point to several microscopic properties that could be driving these improvements. This study verifies improved performance for an industry-ready process and provides insights into understanding the sources of TLS at Si interfaces.

\section{Results and Discussion}

\begin{figure}
    \centering
    \includegraphics[width=\textwidth]{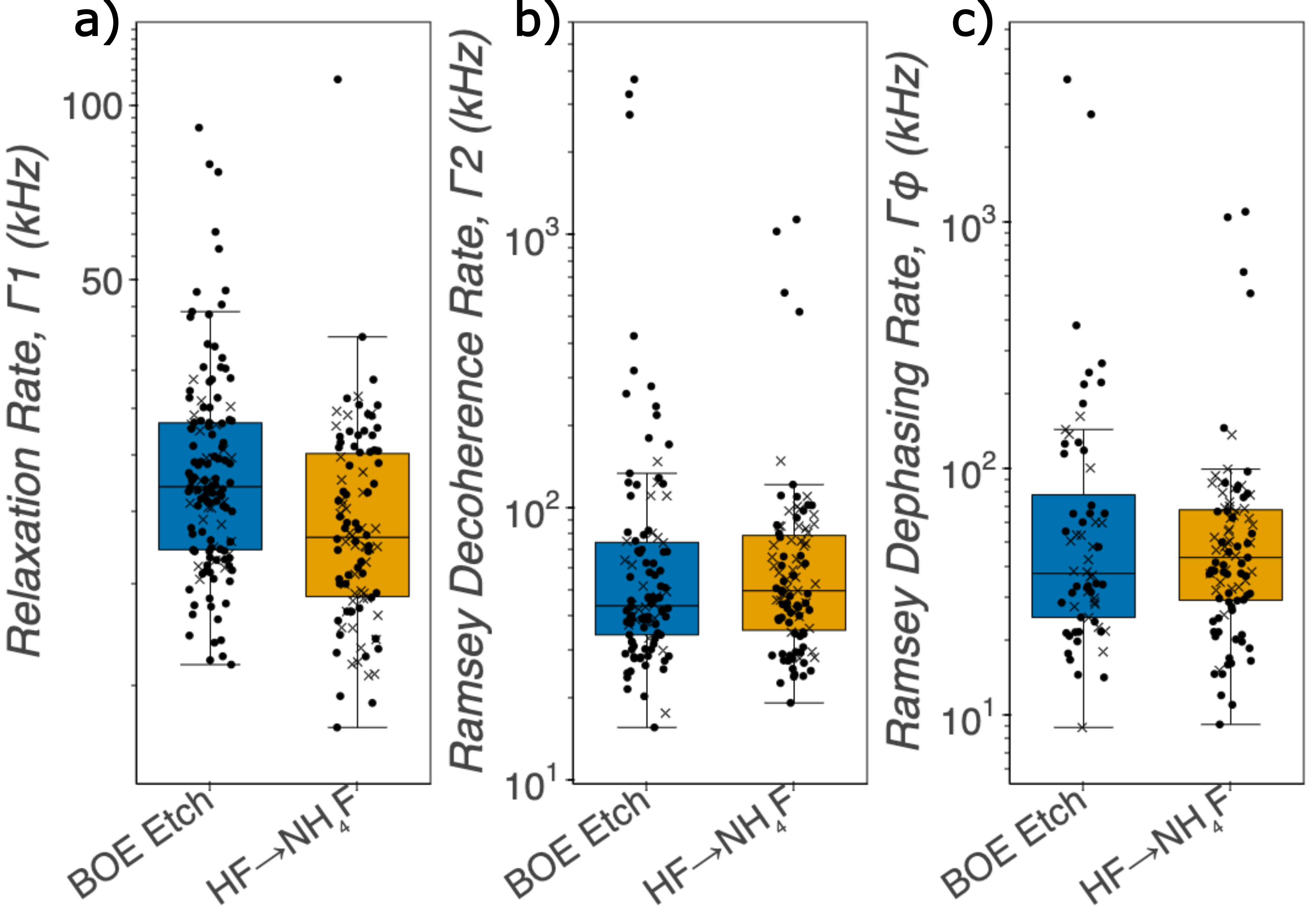}
    \caption{Comparison of measured a) relaxation rate, b) Ramsey decoherence rate, and c) Ramsey dephasing rate for surfaces prepared with a $\ce{BOE}$ etch (blue, left in each set) and $\ce{ HF \rightarrow NH_4F }$ (orange, right in each set). The $\times$ markers represent qubits with Design 1 (uncoupled qubits), and the dot markers represent qubits with Design 2 (coupled qubits). The y-axes are log scale to show the wide range in measured rates (this visually compresses differences between sample set means). }
    \label{fig:t1}
\end{figure}

The dominant sources of loss in superconducting qubits are generally concentrated at interfaces, but specific microscopic sources are numerous and difficult to pinpoint. Given this challenge to understanding which defects at the silicon-air and silicon-aluminum interfaces are affecting performance in these devices, we use qubit performance as the primary metric to evaluate these etch processes. While qubit performance metrics should be very sensitive to changes at these interfaces, we expect some spread across qubits on a single die, and across different wafers due to fabrication non-uniformities. To help control for these variations, we measured multiple dies from multiple wafers, each processed by the etches described earlier. Qubit performance metrics including decoherence are also impacted by design, which affects coupling to the environment and to other devices. To control for design variations, we verify the impact of these etch processes on two separate qubit chip designs; Design 1 is a simple test chip of nominally isolated qubits, while Design 2 is more complex where multi-qubit coupling is employed. Comparing multiple designs also ensures that any improvements carry over to devices useful for potential applications in quantum computing, and not only in test platforms. Both designs use a similar concentric qubit design, while the layouts, qubit-qubit coupling, and surface participation ratios are different. Design 1 consists of 14 flux tunable qubits and 2 fixed qubits, without any qubit-qubit couplers. Design 2 consists of 32 flux-tunable qubits with fixed couplers to 2 or 3 neighbors in a square-octagon layout (similar to the Aspen-9 lattice described in Li \textit{et al.} \cite{Li_Alam_Iadecola_Jahin_Job_Kurkcuoglu_Li_Orth_ozguler_Perdue_etal._2023}), where we only measure 24 qubits on each die due to test setup limitations. These 32-qubit devices are flip-chip bonded to a superconducting cap for signal delivery and isolation, and as a result are expected to have different surface participation ratios from the simpler device without a cap. The results in Fig. \ref{fig:t1} show decoherence values for both device designs, and confirm that the results are consistent across both die layouts. 

For each set and design of devices, we measured multiple chips from the same wafer, as well as different wafers for each design type, and qubit measurements show that $\ce{ HF\rightarrow NH_4F }$ etched devices have a median relaxation rate $\Gamma_1=18 \ \text{kHz}$ with 1st and 3rd quartiles $15,\ 25\ \text {kHz}$ (corresponding to a median $T_1 \approx 55\ \mu s$), while the $\ce{BOE}$ etch set of devices have a median $ \Gamma_1=22 \ \text{kHz} $ with 1st and 3rd quartiles $17,\ 28\ \text{kHz}$ (corresponding to a median $T_1 \approx 45\ \mu s $). The Tukey HSD (honestly significant difference) test reports that these are different distributions with $p=0.002$, and the $95\%$ upper and lower confidence intervals for the difference between two sets are $ 1.8\text{ to }8\ \text{kHz}$. We report the median and quartiles here to avoid skewing the results with either high- or low-performing outliers. Despite the wide spread in each set, we can confidently conclude that these distributions are different, and the $\ce{HF\rightarrow NH_4F}$ set has improved relaxation rate (and $T_1$). We do not find any significant differences between the sets for $\Gamma_2$ or $\Gamma_\phi$, and since we note that the decoherence and dephasing rates are higher than the relaxation rate, we believe that $\Gamma_2$ and $\Gamma_\phi$ are limited by external factors (i.e., not limited by relaxation). The same trend holds when comparing coherence results for each chip design individually.

\begin{table}
    \centering
    \begin{tabular}{ccccc} 
    \hline
 Group& Tunable Qubits& Sum of Qubit Tunable Ranges& Number of TLS&SPI\\ 
 \hline
         $\ce{BOE}$& $73$& $38.3\ GHz$& $73$& $0.0095$\\ 
         $\ce{HF\rightarrow NH_4F} $& $95$& $63.1\ GHz$& $68$& $0.0065$\\ 
         \hline
    \end{tabular}
    \caption{Summary of the strongly-coupled TLS observed in qubits from both sets}
    \label{tab:TLS}
    
\end{table}
We compare the effects of strongly-coupled TLS in these sample sets by spectroscopically measuring the qubit frequency as a function of flux bias, and fitting any features or avoided crossings to measure the coupling strength. The splitting in the avoided crossing is given by $\sqrt{\Delta^2_{TLS}+4g^2}$, where $g$ is the coupling between the TLS and the qubit, and $\Delta_{TLS}$ is the difference in frequency between the qubit and the TLS. We find that the $\ce{ HF \rightarrow NH_4F}$ etched devices are improved by having both fewer TLS, and TLS with smaller coupling strength. In terms of number of TLS per tunable range, there are on average $\sim1.1\ \text{TLS/GHz}$ in the $\ce{ HF \rightarrow NH_4F }$ set compared to $\sim1.9\ \text{TLS/GHz}$ in the $\ce{BOE}$ etched set. We also use the metric SPI (Spectral Pollution Index) to track the frequency-normalized effect of strongly-coupled TLS across qubits in an experiment where the qubits may have different tunabilities. SPI is the ratio of the sum of all TLS coupling strengths in a set of qubits, normalized by the sum of all the tunable qubit frequency ranges (fmax-fmin). We find that the SPI (the ratio of TLS coupling strengths to the tunable range) is reduced in the $\ce{ HF \rightarrow NH_4F }$ set by about 32\%, as shown in Table \ref{tab:TLS}.
 
\begin{figure}
    \centering
    \includegraphics[width=\textwidth]{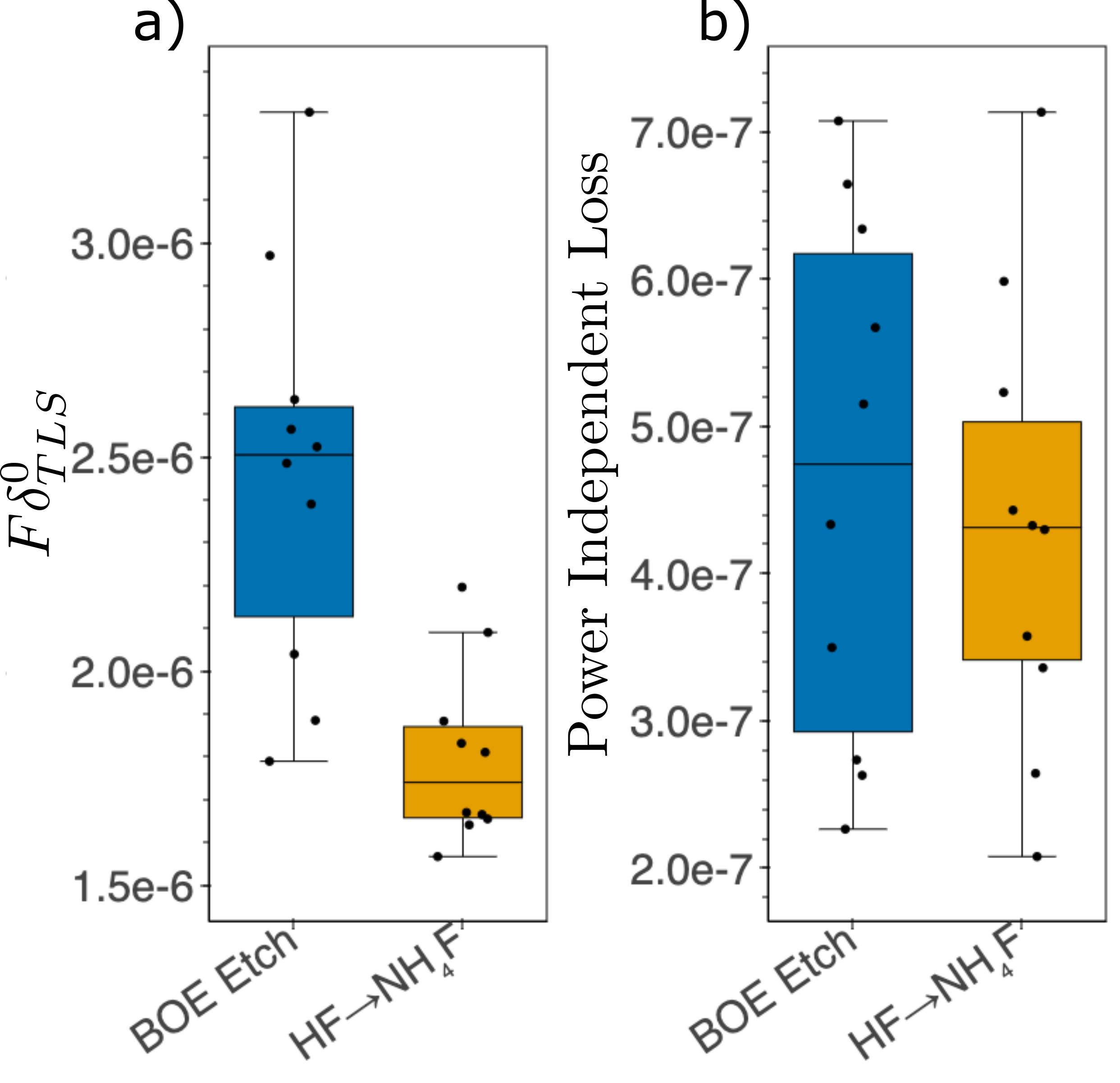}
    \caption{ a) TLS loss and b) power-independent losses extracted from power-dependence measurements on coplanar waveguide resonators fabricated with both etch methods.}
    \label{fig:resonators}
\end{figure}

In addition to qubit measurements, and to help separate the contribution of losses at the Si-air interface from those at Si-Al interface, we measure the power-dependent loss tangent of ten coplanar waveguide resonators for each method. We find that there is a significant difference in losses due to TLSs, with the $\ce{BOE}$ and $\ce{ HF \rightarrow NH_4F }$ treated resonators having $F \delta^0_{TLS} = 2.5 \times 10^{-6}$ and $1.7 \times 10^{-6}$ respectively, representing 32\%\ lower TLS losses in the $\ce{ HF \rightarrow NH_4F }$ set, as shown in Fig. \ref{fig:resonators}(a), consistent with the reduction in SPI measured on the qubits. The power-independent losses for the two etch processes, however, are similar (within 8.5\% of each other, with a large spread), with $\delta_{PI} = 4.7 \times 10^{-7}$ and $4.3 \times 10^{-7}$ for $\ce{BOE}$ and $\ce{ HF \rightarrow NH_4F }$ sets respectively, as shown in Fig \ref{fig:resonators}(b). Since the power-independent losses are similar between the two sets, we conclude that there is no difference in loss due to hydrogen incorporation in the films \cite{Torres-Castanedo_Goronzy_Pham_McFadden_Materise_MasihDas_Cheng_Lebedev_Ribet_Walker}. Based on the $\ce{NbO_x}$ etch rates described in \cite{altoe_localization_2020} we expect that the short etch times are insufficient to significantly remove the Nb-surface oxide, and so we do not expect any differences in loss from the Nb-air interface. Additionally, we performed TOF-SIMS depth profiles on the niobium layers of resonators treated with different etches and do not find any significant differences in the degree of H incorporation within the Nb film for the two etch methods, nor do we see differences in other impurity concentrations including O, C and F. (see Fig. \ref{fig:SIMS-Si} for the results).

\begin{figure}
    \centering
    \includegraphics[width=\textwidth]{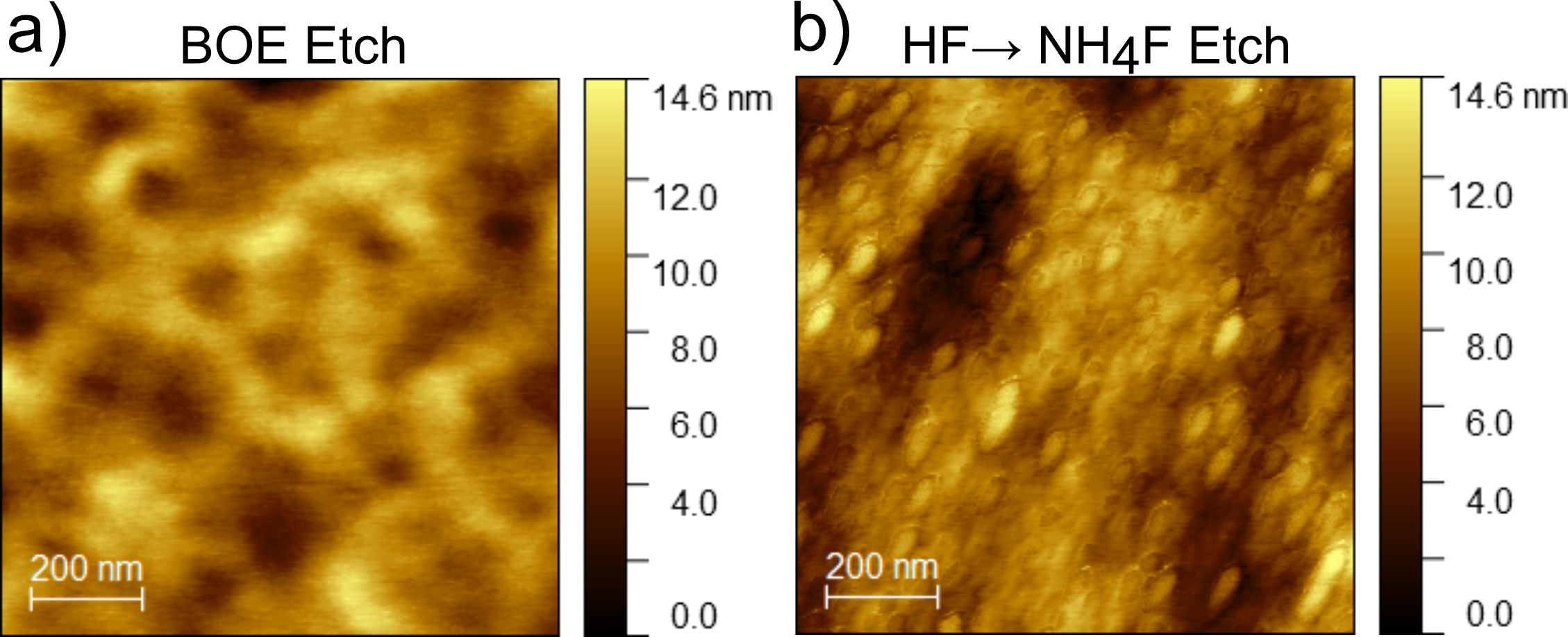}
    \caption{ Surface topology of a $1\ \mu m \times 1\ \mu m$ area measured by AFM on the Si-area near a JJ within a qubit area. The Si surface was etched first with RIE then with the wet chemical process a) $\ce{BOE}$ etch with RMS Roughness $1.95\ \text{nm}$, and b) the $\ce{HF\rightarrow NH_4F}$ etch treatment with RMS roughness $2.29\ \text{nm}$. Larger area scans are in the supplement Fig. \ref{fig:SI-afm-wide}.} 
    \label{fig:AFM}
    
\end{figure}

Since this short etch process is performed before the Josephson junctions deposition, the most likely location for any microscopic changes that could explain the reduced loss are at the Si surface (Si-air interface, or the Si-Al interface underneath the Josephson junction's lower lead). We looked into potential changes at the Si-air interface by performing the $\ce{BOE}$ and $\ce{ HF \rightarrow NH_4F}$ etches on both RIE-etched Si surfaces near the Josephson junction (real devices with the full qubit process), and on bare Si substrates, without any dry etch processes. In both the real devices (Fig. \ref{fig:AFM}) and on the bare substrates (see Fig. \ref{fig:SI-AFM-bareSi}), we observe that Si substrates etched with $\ce{HF\rightarrow NH_4F}$ show the presence of some small diameter surface peaks $10\ \text{nm}$ high in AFM scans, increasing the roughness over that of the $\ce{BOE}$ etched surfaces. However, the background roughness (presumably from the RIE etch process) is decreased in the $\ce{HF\rightarrow NH_4F}$ etched samples. We hypothesize that the $\ce{HF\rightarrow NH_4F}$ etch process may reduce some of the large-scale roughness, but that small peaks are created during the HF process step in the $\ce{HF\rightarrow NH_4F}$ etch process. This is consistent with the conclusions in Aldinger \textit{et al.} \cite{Aldinger_Hines_2012} and with the observation that Si substrates etched with $\ce{NH_4F}$ only (skipping the $\ce{HF}$ etch) do not show this additional roughness (see Fig. \ref{fig:SI-AFM-bareSi}). 

To further interrogate the Si surfaces, we measured the contact angle, where we also observe differences between the two treatments. The contact angle on the $\ce{ HF \rightarrow NH_4F }$ treated Si shows a higher angle, indicating a more hydrophobic surface. More hydrophobic surfaces may be caused by more H-passivation on the Si surface, which further suggests that the surface has less Si oxide. See data in the supplement Fig. \ref{fig:contact_angle-SI}.

\begin{figure}
    \centering
    \includegraphics[width=\textwidth]{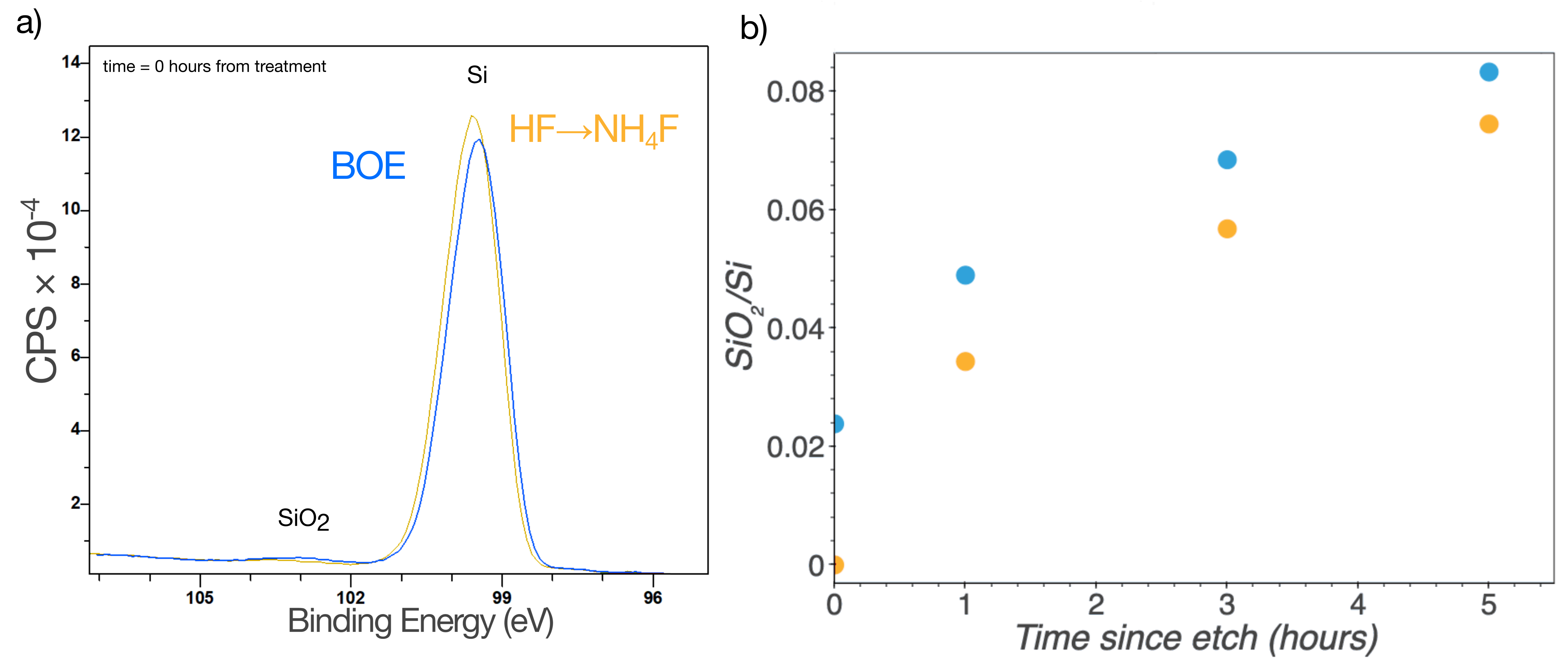}
    \caption{Re-oxidation of Si surfaces in atmosphere after receiving a $\ce{BOE}$ etch or a $\ce{HF \rightarrow NH_4}F$ treatment, as measured by XPS. a) shows an overlay of the measured XPS Spectra at time=0 hours , and b) shows the ratio of $\ce{SiO_2}$ to $\ce{Si}$ peak intensities over time. Uncertainties in the $\ce{SiO_2}/\ce{Si}$ ratio are $<2.5\%$.}
    \label{fig:xps}
\end{figure}

We used X-ray photoelectron spectroscopy (XPS) to examine the chemical nature of the Si surface immediately following treatment by both etch methods and also observe the rate of regrowth of oxide following etching. As shown in Fig. \ref{fig:xps}(a), XPS of the $\ce{Si 2p}$ region taken on Si samples immediately following etching with both methods show a prominent peak at approximately 99.4 eV, consistent with element Si. However, only the BOE treated sample shows signal in the range of 102-104 eV, consistent with the presence of Si oxide. Furthermore, we used XPS to track the growth of the Si oxide signal for up to 5 hours of ambient exposure following etching (See Fig. \ref{fig:xps-si}(b)). We compare the re-oxidation of the Si surface for both etch methods by looking at the ratio of the peak area of the Si oxide signal to the elemental Si signal (Fig. \ref{fig:xps} (b)) and observe that the while the regrowth of $\ce{SiO_x}$ occurs at similar rates for both conditions, the $\ce{HF\rightarrow NH_4F}$ etch treatment provides an initial resistance to oxidation, with less oxide regrowth seen at all time points compared to the BOE treated samples. These results suggest that the $\ce{HF\rightarrow NH_4F}$ etch treatment could be improving the interface between the Josephson junction and the Si substrate by preventing Si oxide re-growth in the short time between surface cleaning and Josephson junction deposition. 

\begin{figure}
    \centering
    \includegraphics[width=1\linewidth]{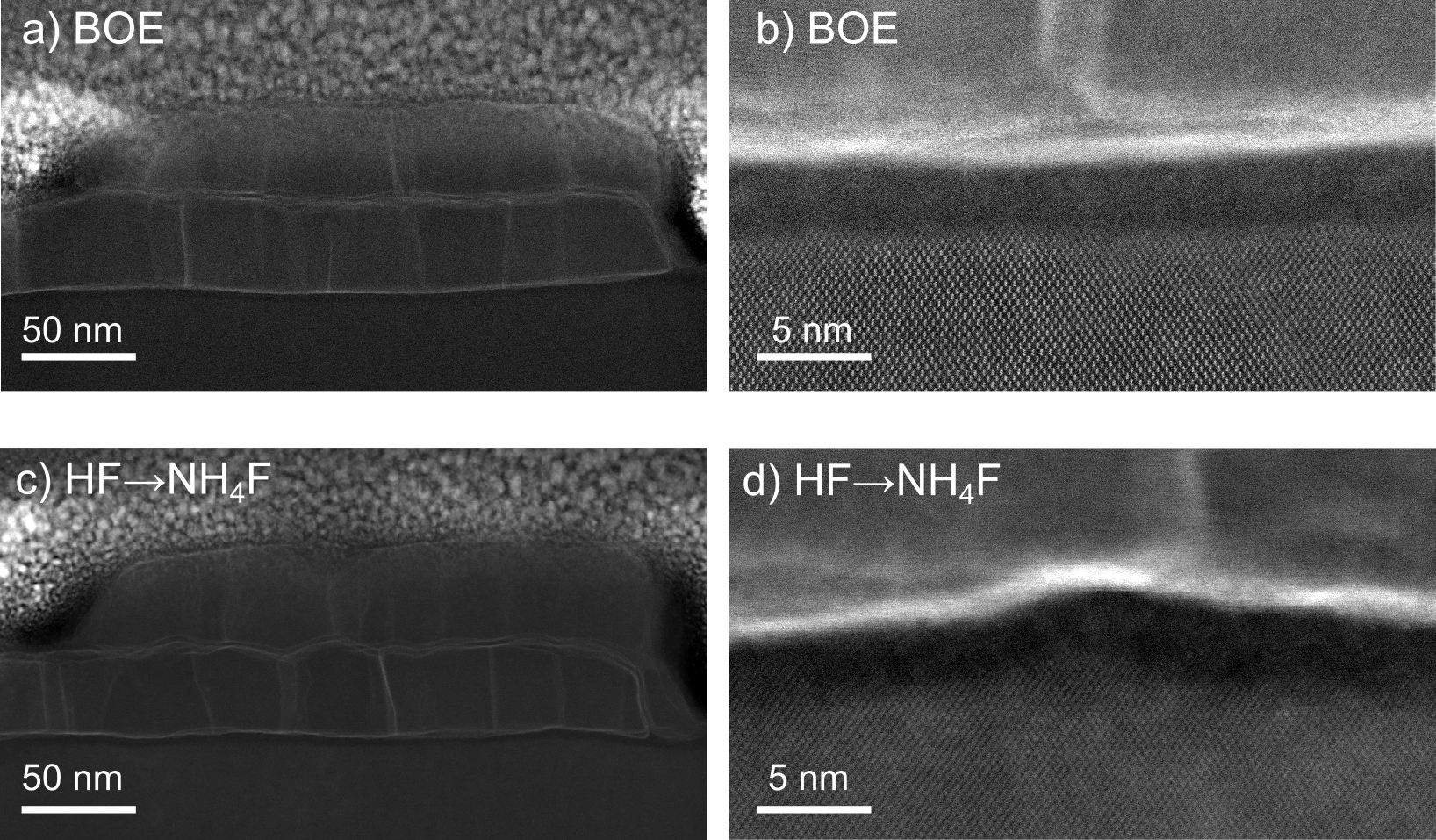}
    \caption{Structural characterization of Josephson junctions. (a,c) Low-magnification STEM images of two representative Josephson junctions with: (a) $\ce{BOE}$ and (c) $\ce{ HF\rightarrow NH_4F }$ substrate cleaning treatments. (b,d) Atomic resolution STEM images showing the amorphous interface between the Si substrate and the Al bottom electrode.
}
    \label{fig:TEM_img}
\end{figure}

To evaluate this hypothesis, we further characterized the interface between the Al bottom electrode and the Si substrate using scanning transmission electron microscopy (STEM) imaging and spectroscopy. Figures \ref{fig:TEM_img}(a-b) display an overview of our Josephson junction by high-angle annular dark field (HAADF) STEM, showing the main components of our device: two polycrystalline Al electrodes with approximately $40-50$ nm thickness, and a thin ($1-2$ nm) $\ce{AlO_x}$ tunnel barrier. The interface between the Al bottom electrode and the Si substrate is shown in atomic-resolution STEM images (Figs. \ref{fig:TEM_img}b,d) for both $\ce{BOE}$ and $\ce{ HF \rightarrow NH_4F }$ treated Si substrates. The Al-Si interface in both cases is amorphous, and its width ranges from $2-4$ nm.

\begin{figure}
    \centering
    \includegraphics[width=1\linewidth]{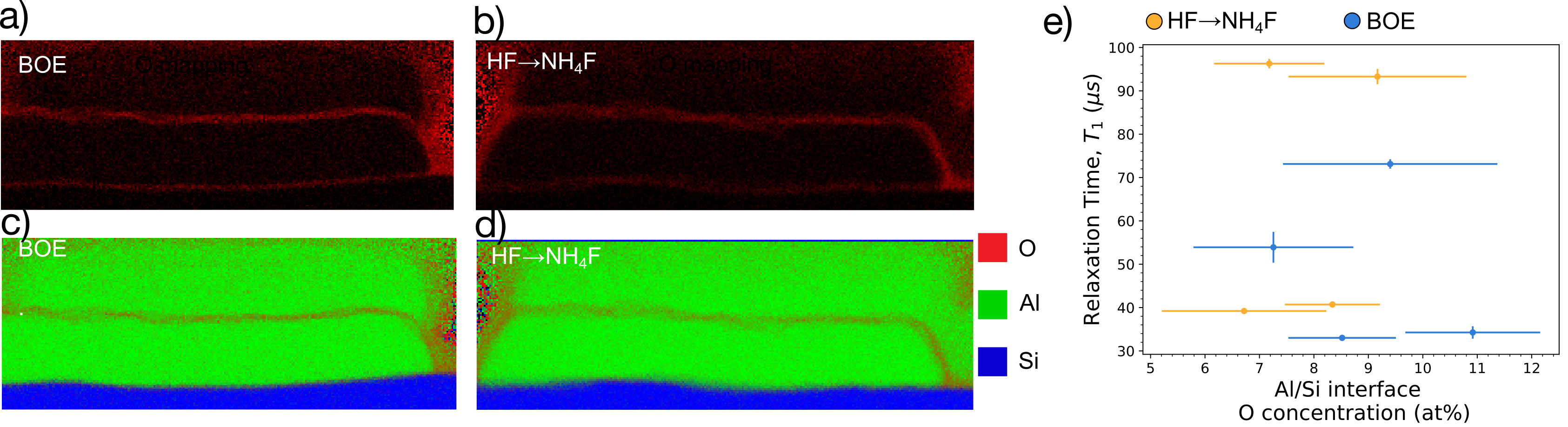}
    \caption{Chemistry of the Al-Si interface. Oxygen chemical mapping and an overlay of all chemical element (O, Al, Si) distribution for $\ce{BOE}$ etched sample (a,c) and $\ce{ HF \rightarrow NH_4F }$ sample (b,d), respectively. (e) Quantification of O concentration at the Al-Si interface for eight devices, four using $\ce{BOE}$ and four using the $\ce{ HF \rightarrow NH_4F }$ method. Error bars in (e) represent the standard error of the mean.
}
    \label{fig:EDS}
\end{figure}

We further employed energy dispersive spectroscopy (EDS) to investigate the chemistry of the aluminum-silicon interfaces. Figure \ref{fig:EDS} shows the chemical distribution of oxygen (O) and of all observed elements (Al, Si, O) displayed in a combined map for a $\ce{BOE}$ etched (Fig. \ref{fig:EDS}(a,c)) and a $\ce{ HF \rightarrow NH_4F }$ etched (Fig. \ref{fig:EDS}(b,d)) device, respectively. As expected, O concentrates at the AlOx barrier between the two Al electrodes. Additionally, EDS mapping shows a significant amount of O at the Al-Si interface for both sample types. We quantify and plot the interfacial O for eight devices, in which four were fabricated with the $\ce{BOE}$ etch treatment, while the other four with the $\ce{ HF \rightarrow NH_4F }$ etch treatment, selecting two high $T_1$ -and two low $T_1$ devices from each set. Overall, the devices that received the $\ce{ HF \rightarrow NH_4F }$ etch show lower O concentration, $7.36\pm 2.66\ \text{at}\%$, compared to the devices that received the $\ce{BOE}$ etch treatment, with $9.54\pm 2.90\ \text{at}\%$ O. However, when we compare the measured $T_1$ of each specific qubit to the EDS-measured average O concentration at the Josephson junction's Al-Si interface, we do not see a clear trend, as shown in Fig. \ref{fig:EDS}(e).

\section{Conclusion}

We show that a chemical change to the surface treatment of the Si substrate, prior to Josephson junction deposition, can result in a significant and reproducible improvement to dissipation properties ($T_1$) and a reduction in the number of TLS per tunable frequency span. We sought to correlate these changes in dissipation and TLS with physical and chemical changes at the silicon-aluminum interface through surface analysis, quantitative electron microscopy, and TOF-SIMS techniques. We observed that qubits treated with $\ce{ HF \rightarrow NH_4F }$ show on average less $\ce{O}$ at the $\ce{Al-Si}$ interface across the population, but on an individual-device basis this did not deterministically predict $T_1$ values. Because we observe a similar reduction in TLS loss in resonator power-dependence and in the strongly-coupled TLS observed in qubit measurements, as well as given that the resonators do not have a substrate-Josephson junction interface, we infer that many TLS may be localized on the Si-air interface, and are not necessarily exclusive to the substrate-Josephson junction interface. This is an unexpected result since the Si-air interface is exposed to uncontrolled atmosphere, forms a native oxide, and the main difference we found at this interface is that the $\ce{ HF \rightarrow NH_4F }$ (with fewer TLS) has additional roughness on the Si surface. TEM imaging of the Si-Josephson junction interface did not observe any significant differences, but this is because TLS losses are likely to be specific localized defects, which are difficult to image or pinpoint with characterization techniques that only analyze a small fraction of the device area. The difficulty in correlating qubit relaxation with microscopic materials properties highlights the significance of performing careful measurements on a large numbers of qubits, compared against a control group before making conclusions on the impacts of fabrication process changes. Despite these challenges, our findings demonstrate that the $\ce{ HF \rightarrow NH_4F }$ etch process improves qubit dissipation and TLS properties and can be easily incorporated into production fabrication processes. 

\section{Acknowledgments}
\noindent This material is based upon work supported by the U.S. Department of Energy, Office of Science, National Quantum Information Science Research Centers, Superconducting Quantum Materials and Systems Center (SQMS) under contract no. DE-AC02-07CH11359. This work made use of the EPIC and Keck-II facilities of the Northwestern University NUANCE Center, which has received support from the SHyNE Resource (NSF ECCS-2025633), the IIN, and the Northwestern MRSEC program (NSF DMR-2308691). This work also made us to the QSET facility of the Northwestern University Center for Applied Physics and Superconducting Technologies (CAPST), which has received support from SQMS.

\printbibliography
\clearpage

\title{Supplement for: Enhanced Superconducting Qubit Performance Through Ammonium Fluoride Etch}
\maketitle

\section{Device Fabrication}
Qubits are fabricated by depositing niobium on $\text{high-}\rho$ Si substrates (slightly n-type, $>10,000 \ \Omega \cdot cm$) and etching the Nb circuit layer with a dry plasma etch. The wafer is cleaned with a wet chemical etch (either  $\ce{BOE}$ or $HF$ followed by $\ce{NH_4F}$ ) before the $Al/AlO_x/Al$ Josephson junctions are fabricated on top of the etched Si surface using a bridgeless double-angle lift-off process, as-described in \cite{Nersisyan_manufacturing_2019} The junctions are connected to the niobium by an aluminum patch (bandage) that is deposited in a subsequent lift-off step. 

\section{Transmission Electron Microscopy}
Transmission Electron Microscopy (TEM) samples were prepared by focused ion beam (FIB) using a dual-beam FEI-SEM Helios Nanolab. The bulk-out and lift-off processes were carried out at 30 kV $\ce{Ga+}$ ion, with the final cleaning step at 5 and 2 kV to remove surface amorphous materials (if any). The final sample thickness is typically 50-100 nm. TEM and scanning transmission electron microscopy (STEM) data were collected on an aberration-corrected JEOL ARM200 S/TEM operating at 200 kV. The convergent angle for ADF-STEM imaging is 25 mrad. EDS data were collected at 200 kV with Dual SSD EDS detector (1.7 sr). Data processing (denoising, background subtraction and signal mapping) was conducted using Gatan GMS software.

\section{Secondary Ion Mass Spectrometry}
\begin{figure}
    \centering
      \renewcommand{\thefigure}{S-\arabic{figure}}
    \includegraphics[width=1\linewidth]{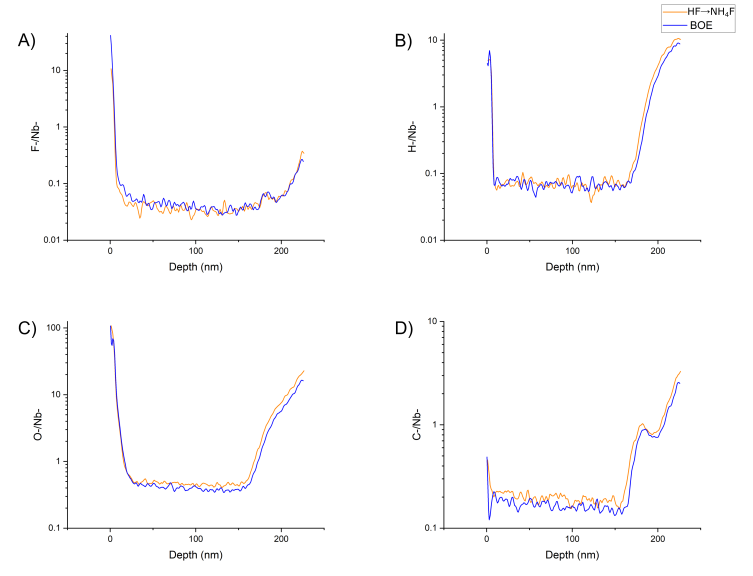}
    \caption{Depth profiles of relative atomic concentration (negative secondary ion yield) through the niobium metal layer obtained using ToF-SIMS. Depth 0 corresponds to the niobium-air surface (this surface was treated with the specified etch process), while depth $\sim165\ \text{nm}$  corresponds to the niobium-silicon interface. A) shows the ratio of Fluorine to niobium, B) shows the ratio of hydrogen to niobium, C) shows the ratio of oxygen to niobium, and D) shows the ratio of carbon to niobium.}
    \label{fig:SIMS-Si}
\end{figure}

Time-of-flight secondary-ion mass spectrometry (ToF-SIMS) was performed on a IONTOF M6 dual-beam system (IONTOF GmbH) using witness resonator samples that received the same processing as the measured devices. The measurements were performed using $\ce{Bi^+}$ ions at 30 keV as the primary ion source with the decector in negative polarity mode. $\ce{Cs^+}$ ions with an energy of 500 eV were used for depth profiling milling a  $200 \times 200\ \mu m^2$ area with a $25 \times 25\ \mu m^2$  central area being the analysis spot. SurfaceLab7 software was used to analyze the SIMS data. We use the ratio of the target ion yield to the niobium matrix yield in order to avoid needing calibrated standards. Depth profiles in Fig. \ref{fig:SIMS-Si} show that the niobium layer does now have any significant differences in $\ce{F^-, H^-, O^-,}$  or $\ce{C^-}$ concentration between the two treatments.  We were unable to perform similar TOF-SIMS analyses on  impurity concentrations at the $\ce{Al-Si}$ interface because the Josephson junctions are smaller than the minimum sample area.

\section{X-ray Photoelectron Spectroscopy}
\begin{figure}
    \renewcommand{\thefigure}{S-\arabic{figure}}
    \centering
    \includegraphics[width=1\linewidth]{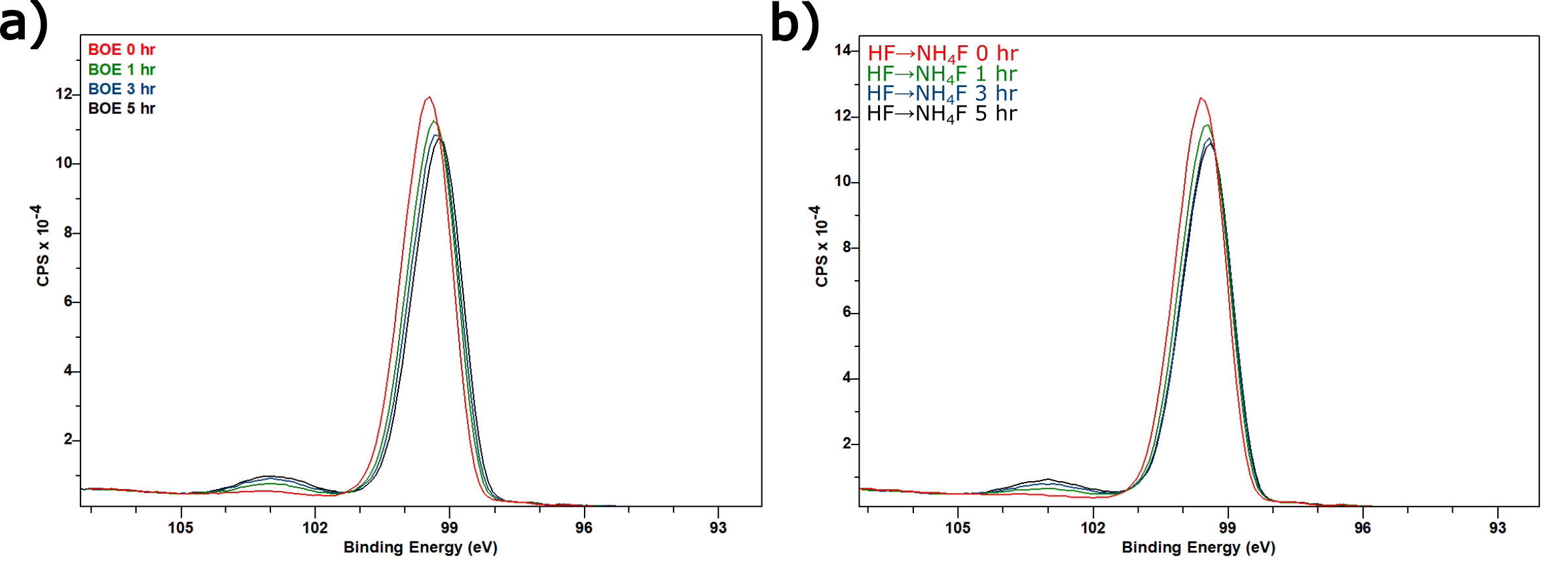}
    \caption{XPS Spectra used to analyze re-oxidation of the Si surface over time for devices treated with a) $\ce{BOE}$ Etch, and b) $\ce{ HF \rightarrow NH_4F }$ etch processes. }
    \label{fig:xps-si}
\end{figure}

X-ray photoelectron spectroscopy (XPS) was performed on a Thermo Scientific ESCALAB 250Xi XPS system equipped with a monochromated $\ce{Al-K_\alpha}$ X-ray source with an energy of 1486.6 eV. The measurement spot size was $\sim 500\ \mu m$ and a flood gun was used for charge compensation. Si surfaces received the specified etch protocol and then were immediately taken to the XPS for measurement, with nominally 15 minutes of air exposure prior to the first time point. Raw results are in Fig. \ref{fig:xps-si}. Samples were stored in ambient conditions for the specified times for subsequent measurements. The data analysis was performed using CasaXPS software. The Si2p orbital was fitted using a U 2 Tougaard background and a Gaussian-Lorentzian peak fit (70\% Gaussian and 30\% Lorentzian). All spectra were charged corrected using carbon (C1s) at 284.8 eV as a reference.

\section{Qubit Measurement Methodology}
For qubit measurements, dies are affixed to a printed circuit board anchored to the mixing chamber plate of a dilution refrigerator with a nominal base temperature of $\approx10\ \text{mK}$ and are housed in superconducting and low-permittivity shielding. Input lines have 86dB of total attenuation, while output lines are isolated by 2 double-junction isolators and amplified by a high electron mobility transistor  (HEMT) at 4 K and low noise amplifiers at room-temperature. Qubits are driven through the readout lines.  Each qubit's coherence properties ( $T_1,\ T_2,\ T_\phi$) are measured many times over $\sim$4 days, and all measurements are performed at $f_{max}$, regardless of whether there is a TLS or neighboring qubit at a similar frequency. In Figure \ref{fig:t1} (in the main text), each data point shown is the median value of all the measurements over the week.  Using the median value avoids any low or high outliers due to single measurement errors. 

\section{Resonator measurements}

Resonator devices treated with both etch methods were measured during the same cooling cycle and shared the input and output lines via cryogenic switches. In order to measure internal quality factors below one photon level, we have reduced the thermal photon population using 70 dB attenuation (20 dB on 4 K stage, 10 dB on still and cold plate stages and 30 dB on the mixing chamber stage) on the input line and Quantum Microwave Components LLC commercial infrared Eccosorb filters on both the input (QMC-CRYOIRF-002) and output (QMC-CRYOIRF-003) ports. The output signal is amplified by a high-electron mobility transistor (HEMT) cryogenic low noise amplifier (LNF-LNC4\_8C) followed by a room-temperature low noise amplifier(LNF-LNR4\_8F), together providing roughly 80 dB of gain. Additionally, at low power, we used a Josephson parametric amplifier (JPA) provided to us by the Advanced Microwave Photonics group at NIST, which amplified the signal at the mixing chamber cryostat stage by additional 10-20 dB of gain with minimal noise addition.

\begin{figure}[htb]
    \renewcommand{\thefigure}{S-\arabic{figure}}
    \centering
    \includegraphics[width=1\linewidth]{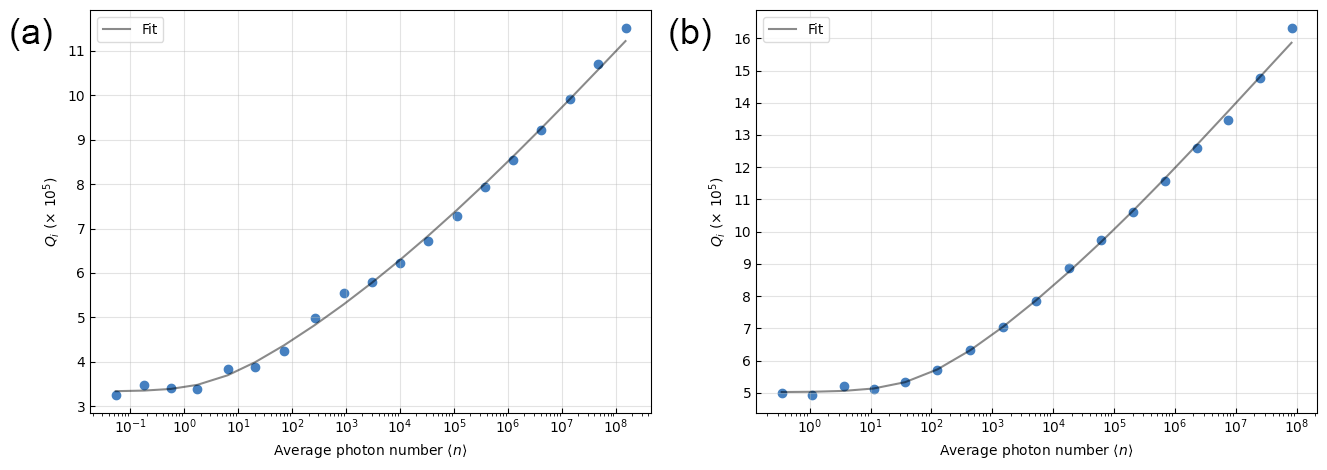}
    \caption{Sample power sweeps of the internal quality factor $Q_i$ for (a) BOE and (b) $\ce{ HF \rightarrow NH_4F }$ resonators.}
    \label{fig:resonator-power-curves}
\end{figure}

For all ten resonators on each device, we performed a power sweep for average photon number $\left\langle n \right\rangle \sim 0.1-10^8$ in order to correctly extract TLS losses information, where Fig. \ref{fig:resonator-power-curves} shows typical curves of the internal quality factor $Q_i$ for the devices. For each resonator, we fit the curve to a TLS loss model \cite{McRae_2020}
\begin{equation}
    \frac{1}{Q_i} = F \delta^0_{TLS} \frac{\tanh{\frac{\hbar \omega}{2 k_B T}}}{\left( 1 + \frac{\left\langle n \right\rangle}{n_c}\right)^\beta} + \delta_{PI},
\end{equation}
where $F$ is the filling factor, $\delta^0_{TLS}$ is the intrinsic TLS loss, $n_c$ is the critical photon number, and $\delta_{PI}$ accounts for power-independent losses. We report the aggregate results in Table \ref{tab:resonators-TLS-params}, where we see a significant difference in the TLS losses, $F \delta^0_{TLS}$, but no difference in the power-independent losses, consistent with the qubit results.

\begin{table}[htb]
 \renewcommand{\thefigure}{S-\arabic{Table}}
\caption{TLS fitting parameters}
\centering
\begin{tabular}{|l|c|c|c|}
\hline
Method & $F \delta^0_{TLS}$ & $n_c$ & $\delta_{PI}$ \\
\hline
$\ce{BOE}$& $(2.5 \pm 0.4) \times 10^{-6}$ & $12 \pm 16$ & $(4.7 \pm 2.3) \times 10^{-7}$ \\
$\ce{ HF \rightarrow NH_4F }$& $(1.7 \pm 0.2) \times 10^{-6}$ & $40 \pm 50$ & $(4.3 \pm 1.8) \times 10^{-7}$ \\
\hline
\end{tabular}
\label{tab:resonators-TLS-params}
\end{table}

\section{Atomic Force Microscopy}
\begin{figure}
    \centering
     \renewcommand{\thefigure}{S-\arabic{figure}}
    \includegraphics[width=1\linewidth]{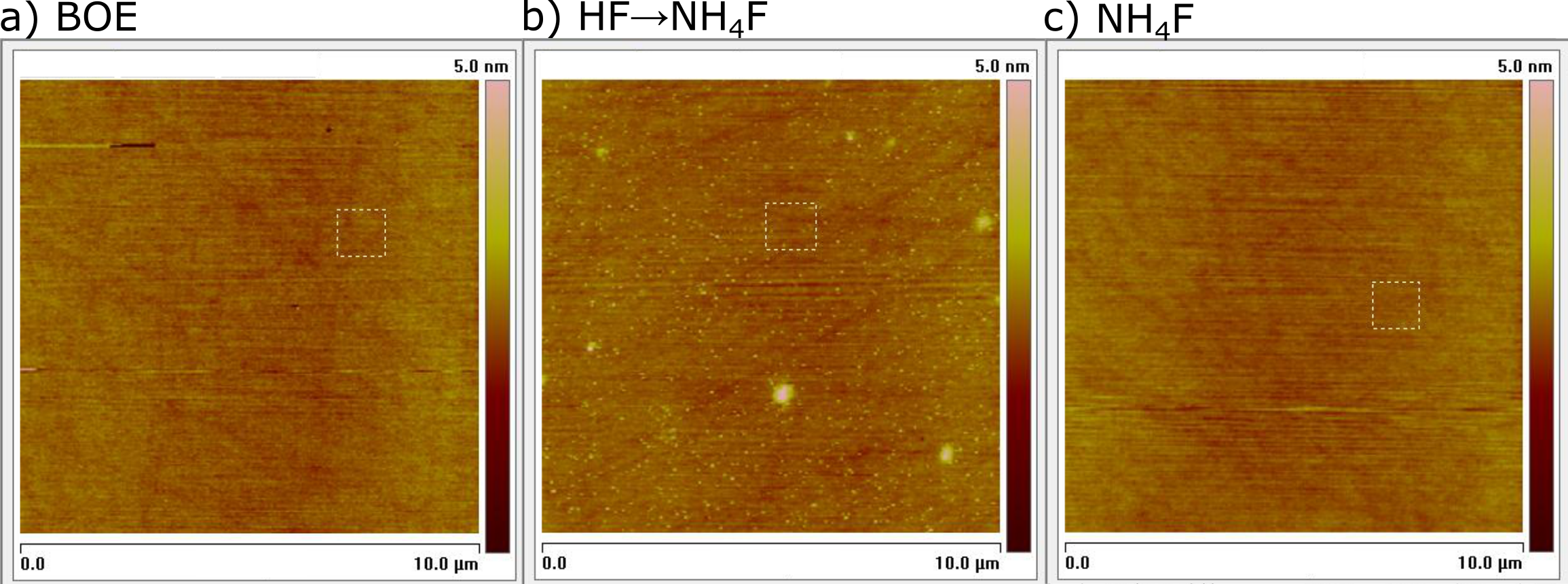}
    \caption{Surface topology of Bare Si substrates treated with each of a) $\ce{BOE}$ etch, b) $\ce{HF\rightarrow NH_4F}$ etch, and c) $\ce{NH_4F}$-only etch treatment.  Each figure shows a $10\ \mu m \times 10\ \mu m$ area measured by AFM. }
    \label{fig:SI-AFM-bareSi}
\end{figure}

\begin{figure}
    \centering
    \includegraphics[width=1\linewidth]{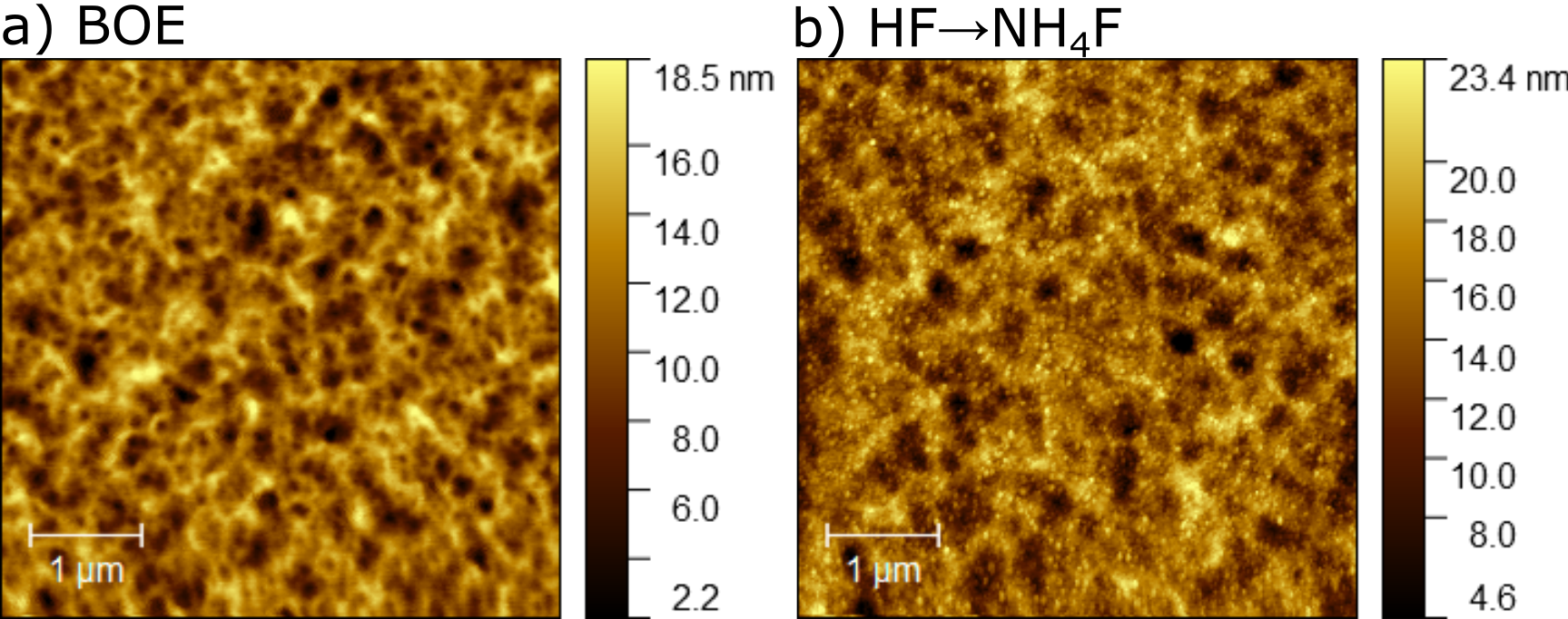}
     \renewcommand{\thefigure}{S-\arabic{figure}}
    \caption{Surface topology of a $5\ \mu m \times 5\ \mu m$ area measured by AFM on the Si-area near a JJ within a qubit area. The Si surface was etched first with RIE then the wet chemical process a) $\ce{BOE}$ etch with RMS Roughness $2.53\ \text{nm}$ over the whole area shown, and b) the $\ce{HF\rightarrow NH_4F}$ etch treatment with RMS roughness $2.98\ \text{nm}$ over the whole area shown.}
    \label{fig:SI-afm-wide}
\end{figure}

Surface roughness is measured on bare substrates (Fig. \ref{fig:SI-AFM-bareSi}) and practical devices (Fig. \ref{fig:SI-afm-wide} ) using Atomic Force Microscopy (AFM).  The figures in the main text are of a smaller area taken from the same samples as Fig. \ref{fig:SI-afm-wide}.  We observe that the $\ce{ HF \rightarrow NH_4F }$ show additional small roughness peaks, that we attribute to the $\ce{HF}$ etch step since these do not appear when treating a bare substrate with $\ce{ NH_4F }$ alone. 

\section{Contact Angle Measurements}
\begin{figure}
    \renewcommand{\thefigure}{S\arabic{figure}}
    \centering
     \renewcommand{\thefigure}{S-\arabic{figure}}
    \includegraphics[width=0.5\linewidth]{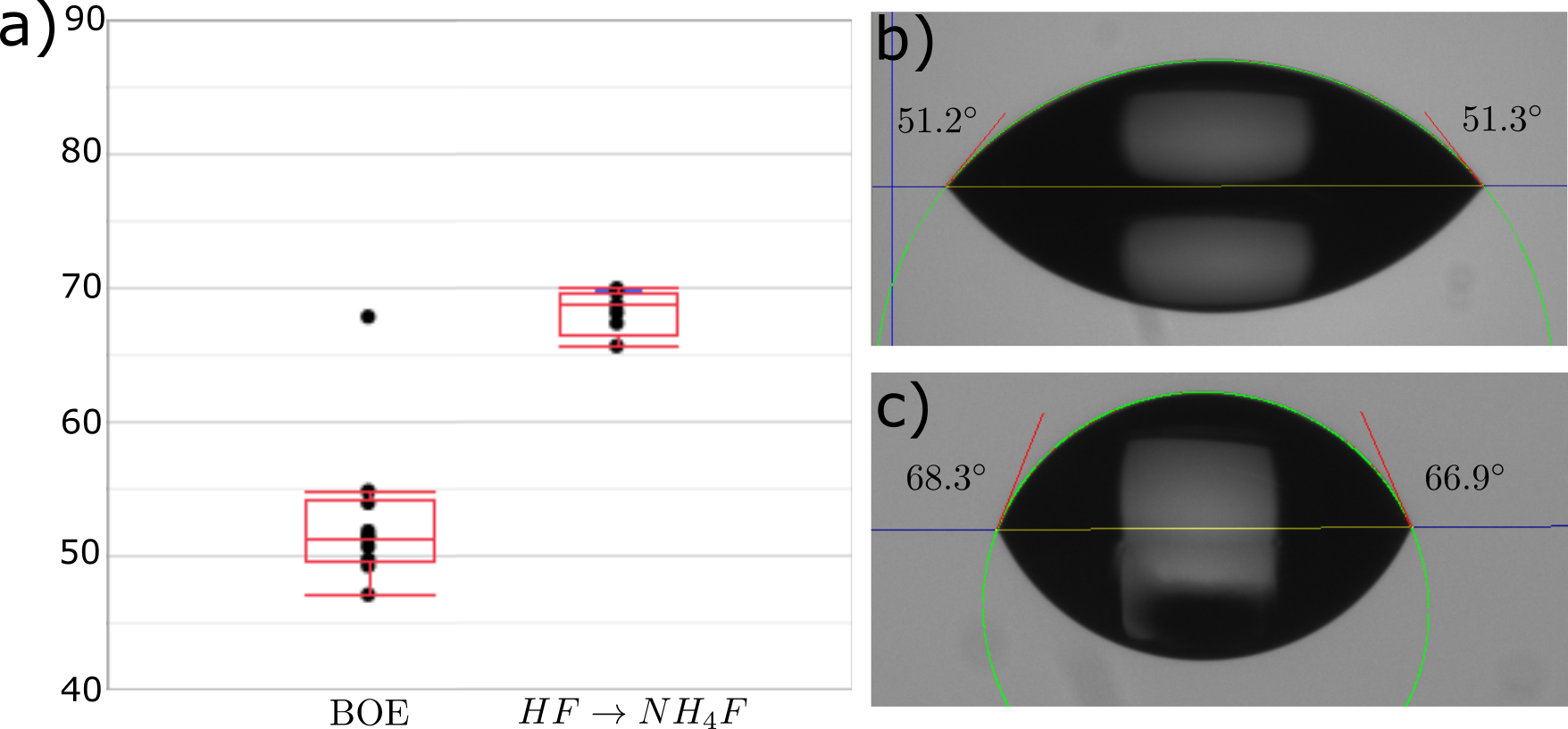}
 \caption{Results of contact angle measurements comparing bare Si wafer surface after treatment with $\ce{BOE}$  and  $\ce{HF\rightarrow NH_4F} $. a) box plot showing distribution of contact angles across multiple locations on a wafer b) Representative example image showing droplet and measured contact angles for $\ce{BOE}$ etched Si. c) Representative example image showing droplet and measured contact angles for $\ce{ HF \rightarrow NH_4F }$ etched Si.}
    \label{fig:contact_angle-SI}
\end{figure}

Contact angle measurements (shown in Fig. \ref{fig:contact_angle-SI} were done to evaluate the wettability of the substrate surface after acid cleaning. Contact angle is the angle at the three-phase boundary (liquid, air, substrate). It is measured by imaging water droplets on the substrate surface. A motorized syringe was used to dispense 1.5 uL droplet of DI water on the surface and the drop shape is then analyzed with dedicated software.

\end{document}